\documentclass[letter,structabstract]{aa} 

\usepackage{graphicx}
\usepackage{txfonts}
\usepackage{natbib}

\bibpunct{(}{)}{;}{a}{}{,}

\begin{document}
   \title{Global e-VLBI observations of the gamma-ray narrow line Seyfert~1 PMN J0948+0022}

\author{
M.~Giroletti\inst{1} \and 
Z.~Paragi\inst{2,3} \and 
H.~Bignall\inst{4} \and 
A.~Doi\inst{5} \and 
L.~Foschini\inst{6} \and 
K.~\'E.~Gab\'anyi\inst{7,3} \and
C.~Reynolds\inst{4} \and
J.~Blanchard\inst{8} \and 
R.~M.~Campbell\inst{2} \and
F.~Colomer\inst{9} \and 
X.~Hong\inst{10} \and 
M.~Kadler\inst{11,12,13} \and 
M.~Kino\inst{14} \and 
H.~J.~van~Langevelde\inst{2,15} \and 
H.~Nagai\inst{5} \and 
C.~Phillips\inst{16} \and 
M.~Sekido\inst{17} \and 
A.~Szomoru\inst{2} \and 
A.~K.~Tzioumis\inst{16} 
}

\institute{INAF Istituto di Radioastronomia, via Gobetti 101, I-40129 Bologna, Italy
              \email{giroletti@ira.inaf.it}
         \and
         Joint Institute for VLBI in Europe, Postbus 2, 7990 AA Dwingeloo, The Netherlands \and
         MTA Research Group for Physical Geodesy and Geodynamics, P.O. Box 91, H-1521 Budapest, Hungary \and
         ICRAR/Curtin Institute of Radio Astronomy, Curtin University, GPO Box
U1987, Perth, WA 6845, Australia \and
         Institute of Space and Astronautical Science, JAXA, 3-1-1 Yoshinodai, Sagamihara, Kanagawa 229-8510, Japan \and
         INAF Osservatorio Astronomico di Brera, I-23807 Merate, Italy \and
         FOMI Satellite Geodetic Observatory, Budapest, P.O. Box 585, 1592 Hungary \and
         Department of Physics, University of Tasmania, Hobart Tasmania 7001, Australia \and
         Observatorio Astron\'omico Nacional, E-28014 Madrid, Spain \and
         Shanghai Astronomical Observatory, Shanghai 200030, China \and
  Dr. Remeis Sternwarte \& ECAP, Universit\"at Erlangen-N\"urnberg, Sternwartstrasse 7, 96049 Bamberg, Germany \and
         Center for Research and Exploration in Space Science and Technology, NASA GSFC, Greenbelt, MD 20771, USA \and
         Universities Space Research Association, Columbia, MD 21044, USA \and
         National Astronomical Observatory of Japan, 2-21-1 Osawa, Mitaka, Tokyo, 181-8588, Japan \and
         Leiden Observatory, Leiden University, P.O. Box 9513, NL-2300 RA Leiden, the Netherlands \and
         CSIRO Astronomy and Space Science, PO Box 76, Epping, NSW 1710, Australia \and
         NIICT, Kashima Space Research Center, 893-1, Hirai, Kashima, Ibaraki, 314, Japan
             }

   \date{Received ; accepted }

  \abstract
   {There is growing evidence of relativistic jets in radio-loud narrow-line Seyfert 1 (RL-NLS1) galaxies.}
   {We constrain the observational properties of the radio emission in the first RL-NLS1 galaxy ever detected in gamma-rays, PMN J0948+0022, i.e., its flux density and structure in total intensity and in polarization, its compactness, and variability.}
   {We performed three real-time e-VLBI observations of PMN J0948+0022 at 22 GHz, using a global array including telescopes in Europe, East Asia, and Australia. These are the first e-VLBI science observations ever carried out with a global array, reaching a maximum baseline length of 12\,458\,km. The observations were part of a large multiwavelength campaign in 2009.}
   {The source is detected at all three epochs. The structure is dominated by a bright component, more compact than $55\,\mu$as, with a fainter component at a position angle $\theta\sim 35^\circ$. Relativistic beaming is required by the observed brightness temperature of $3.4\times10^{11}$\,K. Polarization is detected at a level of about 1\%.}
   {The parameters derived by the VLBI observations, in addition to the broad-band properties, confirm that PMN J0948+0022 is similar to flat spectrum radio quasars. Global e-VLBI is a reliable and promising technique for future studies.}


   \maketitle

\section{Introduction}

Radio-loud (RL) narrow-line Seyfert 1 (NLS1) active galactic nuclei (AGNs) have
received considerable attention in recent years, since they pose a challenge to
current unified schemes \citep{Urry1995}. NLS1 AGNs are characterized by an
optical spectrum with narrow permitted lines FWHM(H$\beta$) $< 2000$ km/s, a
ratio of the equivalent widths of [OIII]$\lambda$5007 to H$\beta$ smaller than
3, and a bump caused by FeII \citep[see, e.g.,][for a review]{Pogge2000}. They
also exhibit prominent soft X-ray excesses \citep{Boller1996,Grupe2004}. These
properties are indicative of very high (near Eddington) accretion rates and
relatively low black hole masses ($10^6 - 10^8 M_\odot$).  Only a small
percentage ($\sim 7\%$) of NLS1 are radio-loud.  In these cases, their flat
radio spectra and VLBI variability suggest that several of them could host
relativistic jets \citep{Zhou2003,Komossa2006,Doi2006}.

\begin{table*}
\begin{minipage}[t]{\textwidth}
\caption{Log of observations. The two-letter station abbreviations are: 
AT -- Australia Telescope Compact Array (ATCA, 6$\times$22m), Cm -- Cambridge (32m, limited to 128 Mbps in all experiments),
Da -- Darnhall (25m, limited to 128 Mbps in all experiments), Ef -- Effelsberg (100m), Ho -- Hobart (26m, limited to 128 Mbps and a single polarization in all experiments), Jb1 -- Jodrell Bank (Lovell Telescope, 76m), 
Jb2 -- Jodrell Bank (MarkII telescope, 32$\times$25m), Ks -- Kashima (34m, limited to 128 Mbps and a single polarization in all experiments), Mc -- Medicina (32m), Mh -- Metsahovi (14m), 
Mp -- Mopra (22m), On -- Onsala (20m), Pa -- Parkes (64m), 
Sh -- Shanghai (25m), Tr -- Torun (32m), Ys -- Yebes (40m), 
Wb -- Westerbork Synthesis Radio Telescope (WSRT, 12$\times$25m)
}
\label{t.log}
\centering
\renewcommand{\footnoterule}{}  
\begin{tabular}{lllllll}
\hline \hline
Epoch \# & Day of  & Freq.     &  Participating telescopes & Longest baseline & beam FWHM & S \\
 & 2009     & (GHz) & & (km) & (mas $\times$ mas, $^\circ$) & (mJy) \\
\hline
0 & Apr 21 & 1.6  & Ef, Cm(+Da), Jb1, Mc, On, Tr, Wb & Jb-Tr, 1\,388 & $35.4 \times 22.9, 11.7$& $173\pm20$ \\
1 & May 23 & 22.2 & Sh, Mp, Ho, Ks, Ef, Mc, On, Cm, Jb2 & Ef-Ho, 12\,359& $0.38\times0.15, 39.1$& $665\pm130$\\
2 & Jun 10 & 22.2 & Sh, Mp, Ho, Ks, Ef, Mc, On, Mh, Ys, Cm, Jb2 & Ys-Mp, 12\,458 &  $0.43\times0.14, 43.3$& $345\pm70$\\
3 & Jul 04 & 22.2 & Sh, Ks, Mp, Ho, Pa, AT, Ef, Mc, On, Mh,  Cm, Jb2 & Ks-Mc, 8\,811& $0.42\times0.35, 39.5$& $445\pm90$\\
\hline
\end{tabular}
\end{minipage}
\end{table*}

   \begin{figure*}
   \centering
   \includegraphics[width=0.9\textwidth,clip]{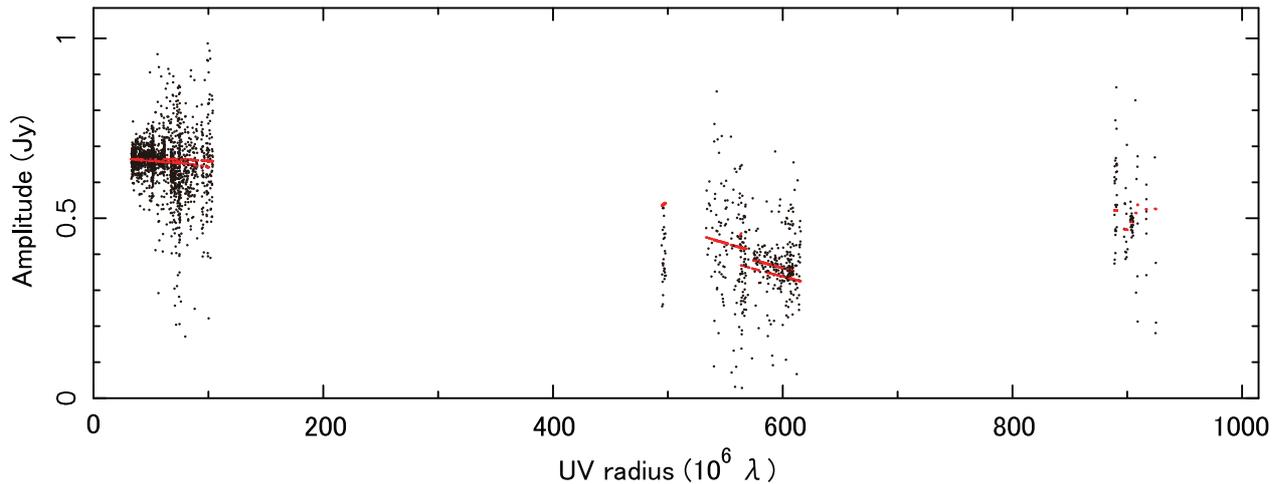}
      \caption{Visibility amplitude versus $(u,v)$-radius for the observations on 2010 May 23. Stokes 'LL' is shown, since Hobart only observed at LCP. The red points show the amplitude visibility of a Gaussian component with flux $S=665\,$mJy and FWHM $0.205 \times 0.055\,$mas in PA $53.1^\circ$.}
       \label{f.radplot}%
    \end{figure*}

The detection by Fermi/LAT of gamma-rays from a handful of RL-NLS1 has provided
definitive confirmation of the presence of relativistic jets in these sources
\citep{discovery,sample,Foschini2010}. A multiwavelength (MWL) campaign was
organized in March-July 2009 to investigate the properties of PMN J0948+0022
with simultaneous observations covering gamma-rays, X-rays, UV, optical, and
radio. This campaign has delivered some interesting results on the emission
processes in this $z=0.585$ source \citep{mwl}. The direct observation of
simultaneous variability across the electromagnetic spectrum has confirmed the
identification of the gamma-ray source with the lower energy counterpart. This
coordinated variability is also characteristic of the relativistic jets present
in the central regions of blazars. Overall, the multiwavelength observational
properties of J0948+0022 seem therefore to indicate that the physics of RL-NLS1
and blazars are similar, and in particular a relationship between RL-NLS1 and
the powerful, high accretion, flat-spectrum radio quasars.

In this letter, we present additional details of the very long baseline
interferometry (VLBI) observations carried out during this campaign. Section~2
describes the procedures of the real-time e-VLBI observations
\citep{Szomoru2008}, which were the first application of this technique using a
global array of antennas in an astronomical program. Section~3 presents the
results, and Sect.~4 discusses conclusions and prospects.



\section{Observations and data reduction}

The observations took place on 2009 April 21, May 23, June 10, and July 4 using
a subset of the European VLBI Network (EVN) and the Long Baseline Array (LBA)
in Australia, and the 34\,m Kashima telescope in Japan. For a complete list of
participating telescopes and other observational parameters, we refer to
Table~\ref{t.log}. The initial observations on 21 April were performed at
1.6~GHz for two hours with the EVN only at 512 Mbps data rate. These
observations served as a pilot project designed to assess the feasibility of
the experiment; we refer to them as epoch $0$.

The other three experiments (epochs $1,2,3$) were carried out with a global
array at 22 GHz for maximum resolution, although Australia-Europe baselines
could not be formed at the third epoch because of the availability constraints
of the Australian telescopes. Each observation lasted for about 11 hours and
was preceded by a 2-3 hour preparation and clock-search run for the
Australian-Asian array, and later included an hour of clock-search and
re-referencing for the whole array including the EVN. The fringe-finders used
were 0420$-$014, OJ287, and 1055+018.  Additional nearby calibrators 0736+017
and 0805$-$077 were regularly observed for continuous fringe monitoring; these
later also served as amplitude calibration check sources. The total flux
densities of these compact sources were obtained from single-dish Effelsberg
(2nd epoch) and synthesis-array ATCA (3rd epoch) measurements. The total bit
rate per telescope was 512 Mbps, divided into four 16~MHz sub-bands in both
polarizations.  The data were streamed in real-time from the stations to the
EVN data processor \citep{Schilizzi2001} at JIVE in all experiments. Plots of
the real-time fringes as well as the final coverage of the $(u,v)$-plane are
shown in \citet{Giroletti2010c}. The longest baseline in our array was the
Yebes--Mopra on June 10, reaching 12\,458\,km.

The data were processed using the NRAO Astronomical Image Processing System
(AIPS).  Amplitude calibration was done with the measured gain curves and
system temperatures. For the LBA stations, nominal SEFDs were used. The $T_{\rm
  sys}$ data were corrected for atmospheric opacity assuming standard weather
conditions for a subset of stations that had opacity-free calibration data.
This in general worked well except for the lowest elevations below
10$^\circ$. All the sources were fringe-fit separately, no phase-referencing
being applied. All of the stations produced fringes, but Kashima had setup
problems in the first two experiments and Hobart had a setup problem in the
first part of the first experiment. After fringe-fitting, the station
amplitudes were further adjusted using the calibrator sources. Phase and
amplitude self-calibration were performed in Difmap to produce the final
images.

The phase response of the right- and left-handed polarized signals with
parallactic angle was corrected before fringe-fitting for all stations,
including the new 40m Yebes radio telescope. For this Nasmyth offset mount
antenna, we used the routines of the {\tt 31DEC10} version of AIPS, following
the procedures described in \citet{Dodson2009}; no RR/LL phase difference was
seen on Yebes baselines after this calibration.  The observations had not
originally been designed to analyze full cross-polarizations; however, we tried
to solve for the D-terms using the bright calibrator 1055+018, limiting
ourselves to the European stations at the second epoch (when there were the
largest number of stations). Reasonable solutions were found, which allowed us
to study the source polarized flux. We did not however correct for the
polarization angle because of the unknown R-L phase difference at the reference
antenna.

\section{Results}

   \begin{figure}
   \centering
   \includegraphics[width=0.88\columnwidth]{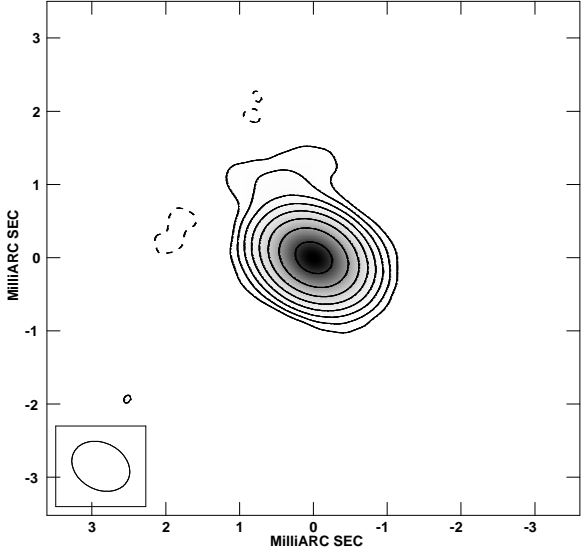}
      \caption{Total intensity contours for J0948+0022 at epoch 2, with levels traced at $(-1, 1, 2, 4, ...)\times1.5$\,mJy\,beam$^{-1}$. The central pixel corresponds to RA$=09^\mathrm{h} 48^\mathrm{m} 57.320^\mathrm{s}$, Dec=$+0^\circ\ 22\arcmin\ 25.560\arcsec$.
              }
         \label{f.total}
   \end{figure}

   \begin{figure}
   \centering
   \includegraphics[width=0.88\columnwidth]{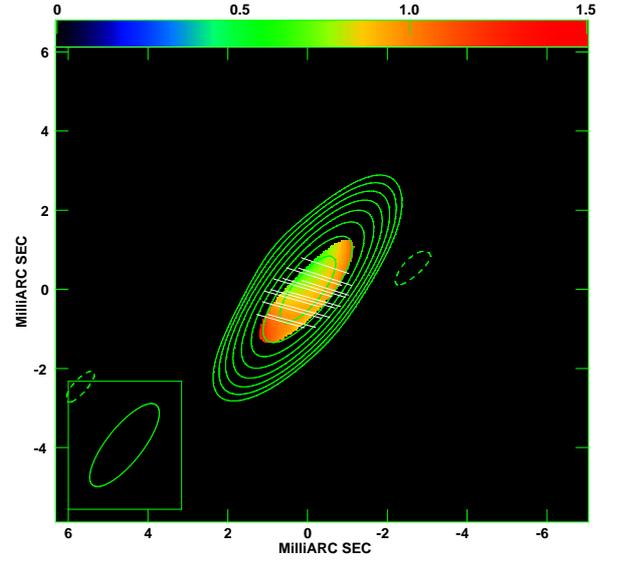}
   \caption{Polarization image from epoch 2, European baselines only. Green contours show total intensity at $(-1, 1, 2, 4, ...)\times1.2$\,mJy\,beam$^{-1}$, with a peak of 250\,mJy\,beam$^{-1}$; white vectors show total polarization (1\,mas\,=\,1.2\,mJy\,beam$^{-1}$, PA not corrected); color shades show fractional polarization (between 0 and 1.5\%).
              \label{f.pol}}%
    \end{figure}

The source is clearly detected at all epochs and frequencies. Inspection of both the visibility plots and the image plane shows that the structure is dominated by a bright compact component. The flux density is difficult to constrain to better than 20\% because of a somewhat incomplete information about the system temperature and gain curve from some stations. It is clear however that there is a flux density of several hundreds milliJanskys revealed at all epochs, with possibly some variation. The overall non-simultaneous spectral index between 1.6 and 22 GHz is clearly inverted, with $\alpha=0.3-0.5$ ($S(\nu)\propto\nu^{\alpha}$). Flux densities from visibility model fitting at all epochs are given in Col.\,7 of Table~\ref{t.log}.

In Fig.~\ref{f.radplot}, we plot the visibility amplitude as a function of $(u,v)$-distance at the first epoch, which is the one with the longest mutual visibility on the longest baselines. 
Despite the uncertainty in the absolute flux level, the relative amplitude scales between telescopes in Europe and Australia are reliably calibrated, as shown by the consistent values across the short $(u,v)$-distances. We are therefore confident that the ratio of the visibility amplitudes on the longest to short baselines is reliably measured, which suggests that that some substructure is present. The fitting of an elliptical Gaussian model to the visibilities yields a component size of $0.205 \times 0.055\,$mas at PA\,$53.1^\circ$. The minor axis corresponds to about one third of the synthesized beam; this means that the source was indeed resolved, thanks to the length of baselines in this intercontinental observation (up to 12\,359\,km between Effelsberg in Europe and Hobart in Australia).

Moreover, the baselines to Shanghai provide visibilities at an intermediate range ($\sim500-600$~M$\lambda$) in NS and EW direction. A smaller visibility amplitude relative to that for the longest baselines indicates that there is structure in the direction transverse to the longest baselines (PA\,$\sim-45^\circ$). In the second epoch, the Fourier transform of the visibility plane, cleaning, and self-calibration are able to more clearly constrain the total intensity structure. A jet component is found at $r=0.86\,$mas from the core in PA $\theta=35.4^\circ$ (see Fig.~\ref{f.total}).
This is also consistent with the lower frequency structure, which shows a resolved jet feature in PA\,$\sim30-40^\circ$ \citep{Doi2006,mwl,Foschini2011}.

The brightness temperature in the source rest-frame based on the model-fit is $\sim3.4\times10^{11}$~K for the first epoch, which is comparable to the equipartition brightness temperature \citep{Readhead1994}. Given the length of the minor axis derived by the model fit, it is conceivable that the size of the core along the major axis is not well-constrained only because of the relatively insufficient baseline lengths.  Therefore, the major axis of the elliptical-Gaussian model component represents only an upper limit, and the resulting brightness temperature is only a lower limit. This confirms that the source must be relativistically beamed, since the observed $T_{\rm b}$ exceeds the equipartition brightness temperature \citep[see][]{Doi2006,Zhou2003}.

Polarization could only be studied for the second epoch and using intra-European baselines. Figure\,\ref{f.pol} provides a polarization map for this epoch. The mean fractional polarization is 0.9\% (peak of 1.3\%), which is barely in excess of the one reported from MOJAVE at 15 GHz at the same epoch. The absolute polarization angle cannot be determined for our data. To our knowledge, this is the first report of a VLBI polarization experiment using the new 40m Yebes radio telescope.

\section{Discussion and conclusions}

\subsection{Scientific implications of the results}

The census of extragalactic gamma-ray sources throughout the EGRET era has been dominated by blazars (flat spectrum radio quasars and BL Lac type objects). The discovery of gamma-ray emission from radio-loud narrow-line Seyfert1 nuclei by {\it Fermi} is therefore of great importance to the study of relativistic jets in AGNs \citep{sample}. First, it has confirmed that these relativistic structures exist in radio loud NLS1, as proposed by \citet{Zhou2003} and \citet{Doi2006}. Moreover, exploiting the sensitivity and the surveying capability of {\it Fermi}, it has triggered the organization of the large MWL campaign including the global e-VLBI observations presented here  \citep{mwl}.

The MWL campaign has revealed variability at all energies, and permitted the estimate of the physical parameters of the jet, such as the dissipation radius ($r=6.7\times 10^{16}$\,cm), the magnetic field ($B=4.1$\,G), and the jet total power \citep{mwl}. A bulk Lorentz factor $\Gamma>1$ was implied by the large Compton dominance; in particular, $\Gamma=10$ was assumed as a typical value. Our lower limit to the brightness temperatures estimated by elliptical Gaussian  model fitting to the global e-VLBI visibilities is a few $\times 10^{11}$~K and provides an independent confirmation that this jet has a Doppler factor greater than one \citep[see also][]{Doi2006,Zhou2003}, as well as in other RL-NLS1 \citep{Gu2010}.  Because of the uncertainty in the overall amplitude scaling, a Doppler factor estimate based on the variability brightness temperature is not  discussed in this study. However, a trend in the total flux density from our observation is present, with a peak measured on 2009 May 23 and a decrease in the following epochs, which are also characteristic features of blazar-like emission. This result  also agrees with other independent variability measurements requiring $\delta>1$ reported at various frequencies in \citet{mwl}.

Significant fractional polarization is also a characteristic of {\it Fermi} detected jets \citep{Hovatta2010}. In J0948+0022 itself,  \citet{Foschini2011} reported an increase in the fractional polarization up to 3\% and a swing of the EVPA of about $90^\circ$ in possible connection with the 2010 gamma-ray outburst. This is similar to what has been reported for other blazars, such as PKS\,1502+106 \citep{Abdo2010}, and might be related to the physical mechanisms responsible for enhanced gamma-ray activity. In this context, our detected value of $\sim 1\%$ fractional polarization seems more characteristic of an intermediate state of activity, consistent with the transition to a low state at high energy observed towards the end of the 2009 MWL campaign \citep{mwl}.

\subsection{Prospects for the use of global e-VLBI}

In this paper, we have presented radio interferometric observations in
real-time e-VLBI mode \citep{Szomoru2008}, using, for the first time, a global
array spanning three continents. The feasibility of transferring data across
continents and correlating them in real time has been demonstrated in a number
of tests and in particular in the LBA observations of SN\,1987A, which were
correlated at JIVE \citep{Tingay2009,Ng2011}. However, these are the first
astronomical e-VLBI science observations on a global scale.

From the very beginning of VLBI, global baselines have been routinely achieved
between e.g., Europe and the US, and some examples exist of truly global
collaborations between arrays on three different continents \citep[see
  e.g.][]{Fomalont2001}.  However, it has always been a challenge to organise
these experiments, especially in the case of time-critical observations such as
our coordinated MWL campaign. The advantages of performing such observations in
real time e-VLBI (when transporting the data through high-speed networks is
possible from all participating telescopes) include the ability to monitor the
network performance in real-time and the possibility of delivering prompt
results \citep{Giroletti2010a,Giroletti2010b}. But most importantly, with the
standardization of the data formats and transport protocols
\citep{Whitney2009}, it will become much easier to create ad-hoc global
networks at relatively short notice, providing excellent $uv$-coverage and
unique long baselines, as well as greatly increased sensitivity due to the
increase in sustainable data rates and the number of telescopes.

Our pioneering global e-VLBI efforts, in spite of some initial technical
difficulties, have convincingly demonstrated the operational feasibility of
this type of observation. Using global baselines and a high observing
frequency, we have probed the innermost region at or above the self-absorption
frequency, thus obtaining the most valuable information as far as the
comparison to high energy activity is concerned. The overall success of the MWL
campaign and in particular of the global e-VLBI observations presented here is
highly encouraging for the continuation of the synergy between high energy
astrophysics and VLBI.

\begin{acknowledgements}
The authors are grateful to Richard Dodson and Daniele Dallacasa for useful
discussions on the polarization calibration, and to Uwe Bach for providing
calibration data from the 100-m telescope of the MPIfR (Max-Planck-Institut
f\"ur Radioastronomie) at Effelsberg.  This activity is supported by the
European Community Framework Programme 7, Advanced Radio Astronomy in Europe,
grant agreement No. 227290 and by the Italian Ministry of Foreign Affairs
within the Scientific and Technology Cooperation Agreement between Japan and
Italy.  e-VLBI developments in Europe are supported by the EC DG-INFSO funded
Communication Network Developments project 'EXPReS', Contract No. 02662 ({\tt
  http:/$\!$/www.expres-eu.org/}). The European VLBI Network ({\tt
  http:/$\!$/www.evlbi.org/}) is a joint facility of European, Chinese, South
African and other radio astronomy institutes funded by their national research
councils. The Long Baseline Array is part of the Australia Telescope National
Facility which is funded by the Commonwealth of Australia for operation as a
National Facility managed by CSIRO.

\end{acknowledgements}

\end{document}